# Unlocking Cryogenic Energy Storage by Constructing Dipole Glass with Unit-cell-level Polar Disorder

Yangyang Si[1,2,#], Denan Li[3,#], Yijie Li[1,#], Changsheng Chen[4,5,#], Jingxuan Li[1], Chao Zhou[1], Hao Xiong[1], Tianfu Zhang[1,6], Wenjin Liao[1], Zhongqi Ren[1], Huaicheng Yuan[7], Dong Li[1], Jing-Kai Qin[1], Cheng-Yan Xu[1], Ye Zhu[5], Yunlong Tang[8], Sujit Das[9], Jieun Kim[10], Junling Wang[2], Hao Pan[11,*], Fei Li,[12,*] Zhen Chen[4,7,*], Shi Liu[3,*], Zuhuang Chen[1,*]

1. State Key Laboratory of Precision Welding and Joining of Materials and Structures, School of Materials Science and Engineering, Harbin Institute of Technology, Shenzhen, 518055, China
2. Department of Physics, City University of Hong Kong, Hong Kong, 999077, China
3. Key Laboratory for Quantum Materials of Zhejiang Province, Department of Physics, School of Science, Westlake University, Hangzhou, Zhejiang, 310024, China
4. Beijing National Laboratory for Condensed Matter Physics, Institute of Physics, Chinese Academy of Sciences, Beijing, 100190, China
5. Department of Applied Physics, Research Institute for Smart Energy, The Hong Kong Polytechnic University, Kowloon, Hong Kong, 999077, China
6. School of Mathematics and Physics, University of Science and Technology Beijing, Beijing, 100083, China
7. School of Physical Sciences, University of Chinese Academy of Sciences, Beijing, China
8. Shenyang National Laboratory for Materials Science, Institute of Metal Research, Chinese Academy of Sciences, Shenyang, 110016, China
9. Materials Research Centre, Indian Institute of Science, Bangalore 560012, India
10. Department of Materials Science and Engineering, Korea Advanced Institute of Science and Technology, Daejeon, 34141, Republic of Korea
11. State Key Laboratory of Advanced Waterproof Materials, Guangdong Provincial Key Laboratory of Nano-Micro Materials Research, School of Advanced Materials, Shenzhen Graduate School, Peking University, Shenzhen, 518055, China
12. Electronic Materials Research Lab, State Key Laboratory for Mechanical Behavior of Materials and Key Lab of Education Ministry, School of Electronic and Information Engineering, Xi'an Jiaotong University, Xi'an, 710049, China.

[#]These authors contributed equally to this work.

**Corresponding authors:** zuhuang@hit.edu.cn; liushi@westlake.edu.cn; zhen.chen@iphy.ac.cn; panh@pku.edu.cn; ful5@xjtu.edu.cn

**Abstract**

Cryogenic energy storage is vital for frontier technologies including deep-space exploration (<90 K) and quantum computing (≤4 K), yet conventional electrochemical energy systems fail below ~230 K due to frozen ion migration. While relaxor-based dielectric capacitors provide high efficiency at room temperature, the intrinsic freezing/growth of polar nanodomains at extended cryogenic regime limits their applications with deteriorated hysteresis losses. Here, we realize superior cryogenic energy-storage performance by designing unit-cell-level disordered dipole-glass state in $Pb_{0.6}Sr_{0.4}ZrO_3$ thin films with composition near antiferroelectric-paraelectric phase boundary. The antiferroelectric-derived dipole-glass introduces enhanced unit-cell-level complexity of dipole interaction that suppresses long-range ferroelectric order. This enables ultralow-hysteresis operation (efficiency > 88%) down to 4 K, delivering record-high energy density (211 J/cm$^3$) at 9 MV/cm, stability over $10^8$ charge/discharge cycles and microsecond-scale charge/discharge capability. This work establishes a dipole-glass paradigm for cryogenic dielectric capacitors, opening a new avenue to highly-efficient energy-storage systems with broad applications in frontier nanoelectronics.

Energy storage systems operable at cryogenic temperatures form the fundamental infrastructure for extreme-condition endeavors including polar-region activities, deep-space exploration, and quantum computing [1-3]. For instance, lunar and Mars rovers require power systems to remain functional at 90 K during nighttime [2], while superconducting quantum processors necessitate 4 K-compatible power sources within dilution refrigerators [3]. However, conventional electrochemical solutions (*e.g.*, batteries, fuel cells, and electrochemical capacitors) generally fail below 233 K due to frozen chemical reaction kinetics and ionic mobilities [4]. Dielectric capacitors, which rely on a physical energy-storage mechanism, *i.e.*, electric displacement and dipole/polarization switching (without long-range ion migration), become a viable option in such circumstances [5]. Dielectric capacitors exhibit distinct advantages including ultrafast discharge rate (high power density) and robust reliability, rendering them highly promising for energy-storage components within electronic devices and electrical power systems, albeit their energy density and charge-discharge efficiency remain to be enhanced [6].

Among numerous materials for dielectric capacitors [7-11], relaxor ferroelectrics (RFEs) such as $Pb(Mg_{1/3}Nb_{2/3})O_3$-$PbTiO_3$ (PMN-PT) and $BiFeO_3$-$BaTiO_3$-$SrTiO_3$ solid solutions have emerged as leading candidates for achieving high energy density and efficiency at room temperature [9, 10]. The high performance is attributed to the polar nanoregions (PNRs) induced by chemical/structural heterogeneities in RFEs, where weak dipole-dipole interactions and suppressed polarization switching barriers significantly reduce the hysteresis loss while maintaining relatively high polarizability [12, 13]. However, the presence of PNRs in RFEs is confined to a limited temperature range associated with fluctuation between nonpolar paraelectric (PE) and polar ferroelectric (FE) phases [14, 15]. Upon cooling RFEs below a freezing temperature ($T_f$), the PNRs grow in size and interaction strength such that the thermal-driven dipole reorientation becomes frozen (forming FE polar order with field cooling) [13-15], inducing increased polarization switching barrier, slowed switching dynamics with pronounced hysteresis loss, and thus substantial reductions in dischargeable energy and efficiency [16]. This fundamental limitation has confined most RFEs to applications above 200 K, creating a critical technological gap for extreme cryogenic environments [15, 16]. Attempts to lower the $T_f$ in order to expand the applicable temperature range, *e.g.*, by simply increasing the ratio of PE components in RFEs to weaken FE order, would inevitably degrade the dielectric polarizability and consequently, the energy density [17]. Therefore, suppressing the FE order without sacrificing polarizability at low temperatures, a task characterized by *inherent contradictions*, is crucial for extending superior energy performance down to cryogenic conditions.

Here, we challenge prevailing paradigms by proposing that antiferroelectric (AFE)-based dipole glass (DG), which does not transform into FE but remain disordered upon cooling (with additional source of local heterogeneities as compared with RFEs), offers a promising pathway for cryogenic dielectric energy storage [18]. By alloying AFE $PbZrO_3$ (PZO) with PE $SrZrO_3$ (SZO), we engineered a DG state in $Pb_{1-x}Sr_xZrO_3$ (PSZO) epitaxial thin films with composition near the AFE-PE phase boundary. Compared with conventional RFEs, this disordered system exploits two synergistic effects: (i) the AFE component provides low-temperature-enhanced antipolar coupling that can complicate the local dipole interactions with chemical heterogeneity, increase the difficulty to form long-range ordering under cooling, and enhance the polarizability to external field; (ii) PE-induced disorder randomly distributes the ferroelectric and antiferroelectric coupling of dipoles in AFE components, confines the dipole-dipole interactions within unit-cell range, flattens energy landscapes, and reduces hysteresis [19, 20]. This DG state, constituting of antipolar AFE structures and nonpolar PE with unit-cell-level disrupted dipole distribution (excluding polar FE components as in conventional RFEs), holds the promise to resolve the inherent trade-off between hysteresis suppression and polarizability retention.

Through molecular dynamics (MD) simulations, atomic-resolution cryogenic electron ptychography, and electrical characterizations, we confirmed the DG nature of $Pb_{0.6}Sr_{0.4}ZrO_3$ thin films, which preserve randomly oriented dipoles and high polarizability below 100 K (extending down to 4 K). This breakthrough delivers unprecedented cryogenic performance, including record-high energy density of 211 J/cm³ at 77 K; high efficiency ($\eta$) >88% maintained across 4-300 K (in contrast to the typical RFE PMN-PT whose $\eta$ drops from 89% at 300 K to 52% at 77 K and 34% at 4 K), ultrahigh breakdown strength (9 MV/cm), and thermally stable charge/discharge speed at microsecond level. These findings establish AFE-based DG as a robust design principle for extreme-condition energy storage, with immediate implications for cryogenic nanoelectronic devices[1-3].

**Atomic-scale insights into the dipole glass state**

To unravel atomic-scale characteristics of the DG state, large-scale MD simulations were conducted on $Pb_{1-x}Sr_xZrO_3$ (PSZO) solid solutions (see Methods). The zero-field dynamic structures from AFE-, DG-, to PE- states are analyzed by computing the distributions of local $Pb^{2+}$/$Sr^{2+}$ displacements (Figs. 1a-c). For the pure AFE component, PZO, MD simulations produce prototype "↑↑↓↓" antipolar dipoles with periodically opposite $Pb^{2+}$ displacements/polarizations (Fig. 1a), agreeing with previous experimental observations [21]. This AFE-state would undergo a first-order transition into a metastable FE-state (with $Pb^{2+}$ displacements switching to parallel) by overcoming an energy barrier, schematically shown as a quasi-triple-well free-energy landscape (Fig. 1a, upper panel) [22, 23]. On the other hand, SZO is commonly viewed as a PE with linear dielectric (LD) response [24]. The dynamic structure by MD simulation reveals an averagely nonpolar matrix with minimal polarization, giving a single sharp energy potential (Fig. 1c). Further, when alloying the antipolar PZO and nonpolar SZO near the AFE-PE phase boundary derived from MD simulations and experiments (*e.g.,* $Pb_{0.6}Sr_{0.4}ZrO_3$) [25], the solid solution exhibits disordered polar dipoles with relatively high polarization at the unit cell level, which does not transform to FE phase but intriguingly maintains a randomly-oriented DG structure with absence of long-range dipole correlation even at 77 K (Fig. 1b). Such a DG system would exhibit distinctive and distributed dipole switching dynamics, characterized by a flattened energy landscape with numerous shallow metastable minima (Fig. 1b), thereby enhancing the cryogenic dielectric responses.

To evaluate the cryogenic energy storage performances of the PSZO solid solutions, we further simulated the polarization-electric field (*P*-*E*) hysteresis loops up to 9 MV/cm for $x$ values ranging from 0.1 to 0.6 at 77 K by MD simulations, extracting the evolution of energy density ($U_r$) and efficiency ($\eta$) as a function of composition (Fig. 1d). We found that $U_r$ reaches the maximum at $x = 0.4$, while $\eta$ generally increases with $x$. This trend can be understood in terms of the trade-off between two factors. First, with increasing Sr content, the maximum polarization $P_m$ decreases monotonically, lowering the theoretical maximum capability of energy storage (Supplementary Fig. 1-3). Meanwhile, the cation displacement distribution $\sigma_p$ (Supplementary Fig. 4), which reflects the strength of the intrinsic driving force for disordering, suppresses the polarization hysteresis associated with the first-order AFE-FE transition, thereby enhancing the dischargeable energy density and efficiency. As a result of the trade-off, the composition of $Pb_{0.6}Sr_{0.4}ZrO_3$ offers the best energy storage performance based on the MD simulations.

To understand the extraordinary cryogenic stability of the DG state, we conducted MD simulations of the dynamic structures and *P*-*E* loops of the $Pb_{0.6}Sr_{0.4}ZrO_3$ in comparison with those of the typical RFE PMN-PT while considering the substrate clamping from 300 K down to 4 K (Figs. 1e-h). Snapshots from MD simulations confirm that, while the PMN-PT (Fig. 1e) and the $Pb_{0.6}Sr_{0.4}ZrO_3$ (Fig. 1g) show similar unit-cell-level disordering of local polarization at 300 K, PMN-PT freezes into a single-domain FE state at 4 K (Fig. 1f). As a result, the relaxor PMN-PT shows substantially increased remnant polarization ($P_r$) and hysteresis loss as temperature decreases (Figs. 1i-j), signifying the formation of FE domains, consistent with prior experimental reports [15, 16]. In comparison, the $Pb_{0.6}Sr_{0.4}ZrO_3$ retains randomly oriented local dipoles even at 4 K after field poling (Fig. 1h), and maintains a near-zero $P_r$ with negligible hysteresis loss (Figs. 1i-j), highlighting the cryogenic stability of DG state. To quantify the dipole ordering in $Pb_{0.6}Sr_{0.4}ZrO_3$ and PMN-PT at 4 K, the spatial dipole-dipole correlation function, $C(r)$, which provide a direct measure of the length scale of polar order, was extracted (Fig. 1k). The PMN-PT system exhibits a positive and stable $C(r)$ that persists over 30 Å, indicating a strong dipole-dipole correlation in the cryogenic FE phase. In stark contrast, the correlation in the DG $Pb_{0.6}Sr_{0.4}ZrO_3$ system decays rapidly and effectively vanishes within ~10 Å (2~3 unit-cells). This indicates a lack of long-range order and points to an ultra-stable, cryogenically disordered glass state maintained at 4 K.

The fundamental mechanism of the persistent DG state in PSZO can be understood as a result of competing interactions that prevent it from settling into a simple, ordered ground state. This phenomenon is analogous to spin glasses that are classically described by the Edwards-Anderson model, which attributes the glassy behavior to a distribution of competing positive (ferromagnetic) and negative (antiferromagnetic) exchange interactions [26]. To determine if a similar mechanism is at play here, we derived the distribution of dipole-dipole interaction energies from the MD simulations, which is defined as the energy cost required to flip a dipole against its local environment (Fig 1l). For PMN-PT, the interaction energies are almost exclusively positive with a mean value of +1.57 eV/dipole at 4 K. This indicates that breaking the ferroelectric coupling between neighboring dipoles incurs an energy cost, reflecting a strong driving force for a long-range polar order. As a result, the system minimizes its free energy by maintain dipole alignment, which suppresses the chemical heterogeneity-driven disorder and leads to the observed long-range correlation (Figs. 1f and 1k). PSZO, on the contrary, exhibits a broad, nearly symmetric distribution of both positive and negative interaction energies, reflecting the presence of competing ferroelectric and antiferroelectric components. These opposing interactions effectively cancel out each other, leading to a much smaller net interaction energy (with a mean value of +0.33 eV/dipole). This energetic stalemate prevents any single ordered state from dominating, thus creating a cryogenically stable disordered DG state.

**Fabrication of solid-solution films with dipole-glass state**

To examine the theoretical predications, a series of ~160-nm-thick $Pb_{1-x}Sr_xZrO_3$ ($x$ = 0.0-0.6) thin films were epitaxially grown on $SrRuO_3$-buffered $SrTiO_3$ (001) substrates via pulsed laser deposition (Supplementary Fig. 6) (see Methods). Reciprocal space mappings (RSMs) around the pseudocubic (pc) (103)-reflections of the PSZO films were performed to investigate the evolution of AFE order (Figs. 2a-c and Supplementary Fig. 7) [27]. For pure PZO films ($x$ = 0.0), the RSM (Fig. 2a) reveals sharp $1/4\{110\}_{pc}$ superstructure (SS) spots and splitting main Bragg spots, reflecting strong AFE order and corresponding AFE distortion-induced tetragonality [28, 29]. As $x$ increases, the 1/4 SS spots exhibit reduced intensity with a streaking feature along the $\langle 110 \rangle_{pc}$ modulation direction indicating gradually disrupted AFE order (Fig. 2b-c and Supplementary Fig. 7). The normalized intensity of the SS spot ($I_{NS}$), defined as $I_{NS} = I_S / I_{(103)}$ (where $I_S$ and $I_{(103)}$ refer to intensity of the SS spot and corresponding PSZO $(103)_{pc}$-reflection, respectively), is also weakened and becomes indistinguishable from the background noise in the RSMs for $x \geqslant 0.4$, indicating the annihilation of long-range AFE ordering (Figs. 2c-d). Meanwhile, the split main Bragg spots move closer to each other as $x$ increases and merge into one single spot at $x$ = 0.4, suggesting that the tetragonality (*i.e.*, $c/a$ ratio) gradually approaches 1 (Fig. 2d). Those structural evolutions demonstrate the presence of AFE-PE phase boundary near $x$ = 0.4, where the long-range AFE ordering is fully suppressed.

To verify the atomic-scale DG state near the AFE-PE boundary and demonstrate its temperature (in)dependence, multislice electron ptychography (MEP) with unprecedented sub-picometer precision was performed on the $Pb_{0.6}Sr_{0.4}ZrO_3$ thin films not only at room temperature (300 K) but also in the cryogenic condition (95 K) (Supplementary Fig. 8), which reaches the equipment limitation (see Methods). As compared with the ordered antiparallel dipole distribution in PZO (Supplementary Fig. 9), the A-site cation displacement mapping (relative to the center of the four nearest B-site cations) reveals chaotic dipole distribution with absence of regular AFE order or definable domain structure in $Pb_{0.6}Sr_{0.4}ZrO_3$ at room temperature (Figs. 2g-h), which agrees with the RSM evolution [30]. As compared with PNRs in conventional RFEs, where the dynamic polar domains usually reach several nanometers [14-16], the polar fragments in the film show smaller dipole correlation length within unit-cell-level, suggesting the formation of DG state derived from chemical heterogeneity and phase competition between AFE and PE. The temperature-driven DG evolution was further investigated with *in situ* cryogenic STEM down to 95 K. The A-site cation displacement (Fig. 2i) and dipole orientation mappings (Fig. 2j) reveal remarkable cryogenic stability of the DG system, where the dipoles maintain typical glassy feature with unit-cell-level randomness at 95 K, without signs of freezing-driven long-range polar order or domain growth as found in conventional RFEs [16]. To quantitively characterize the cryogenic stability, the statistical distribution of displacement magnitude and rotation angle were analyzed at 300 K and 95 K (Figs. 2e-f). The displacement magnitude distribution remains diffusive and exhibits a gaussian feature with a standard deviation of ~2.5 pm at 95 K, demonstrating negligible change compared to the displacement distribution at 300 K (with a standard deviation of ~2.8 pm) (Fig. 2e). Meanwhile, the orientation of dipoles remains homogeneously distributed in all directions at 95 K and almost identical to those at 300 K (Fig. 2f), featuring the extreme disorder of polarization and cryogenic stability in the DG system. These findings thus validate our design principles and MD simulation predictions.

**Breaking the low-temperature barrier over relaxor**

To exam the low-temperature performance of the DG system, comprehensive (di)electric characterizations were conducted on the $Pb_{0.6}Sr_{0.4}ZrO_3$ films in comparison with the typical RFE PMN-PT films. The dielectric data of PMN-PT show remarkable frequency dispersion and broad dielectric peak over a wide temperature range, signifying the typical relaxor behaviors with a large change of dielectric constant (Fig. 3a). In accordance, the *P-E* loops of PMN-PT, while exhibiting minimal hysteresis at room temperature, undergo degradation with substantial increase of hysteresis as the temperature decreases especially below 150 K (Fig. 3b). Such temperature instability severely hinders the application of relaxors as reliable cryogenic electronics. By contrast, the $Pb_{0.6}Sr_{0.4}ZrO_3$ films show negligible dielectric peak when cooling all the way from 450 to 4 K, with minimal variation of dielectric constant (less than ±13.5 %). In parallel, the *P-E* loops of $Pb_{0.6}Sr_{0.4}ZrO_3$ remain slim and stable (with minimal increase of hysteresis) with decreasing temperature down to 4 K (Fig. 3c). Such cryogenic robustness indicates the absence of static freezing of polar dipoles or long-range order domain formation in the $Pb_{0.6}Sr_{0.4}ZrO_3$ system, making it appealing candidate for cryogenic applications, especially high-efficiency dielectric energy storage.

The efficiency $\eta$ of both the DG PSZO and the RFE PMN-PT are extracted from the *P-E* loops in Figs. 3b-c as a function of temperature (Fig. 3d). For PMN-PT, while the efficiency values are near 90% at room temperature, it decreases substantially with decreasing temperature, remaining 50% below 77 K and only 35% at 4 K. In contrast, for $Pb_{0.6}Sr_{0.4}ZrO_3$, the low-temperature efficiency remains over 80% even down to 4 K. This discrepancy of efficiency between PMN-PT and PSZO can be attributed to the suppression of mesoscale ferroelectric domain formation in the dipole-glass PSZO. To quantitatively evaluate the evolution of ferroelectric domains, the remanent polarization $P_r$ of PMN-PT and $Pb_{0.6}Sr_{0.4}ZrO_3$ are provided in Fig. 3e. The $P_r$ of PMN-PT is highly sensitive and increases dramatically (from 4 to 24 $\mu C/cm^2$) with decreasing temperature from 300 to 4 K, indicating the RFE PMN-PT gradually transforms into a normal ferroelectric state [15, 16]. This temperature-driven mesoscale ferroelectric domain formation merges the polar nano regions and enlarges the polarization hysteresis, which becomes the most fundamental obstacle to employ conventional RFE materials for energy storage at cryogenic temperatures. In contrast, the $P_r$ of PSZO shows negligible increase (< 2.5 μC/cm$^2$) with temperature decreasing down to 4 K, suggesting suppression of ferroelectric domain formation at cryogenic conditions. Such favorable properties over conventional RFEs can be ascribed to the more complicated interactions in the AFE-based DG system. While the PE component (SZO) enhances the chemical and structural heterogeneity to realize dipole disorder at room temperature, the AFE component (PZO) brings in antipolar interaction that is believed to be the key to inhibit the growth of polar order (FE domains) with decreased temperature and to maintain the disordered DG state in the cryogenic regime. In addition, the AFE-PE phase competition flattens the energy barrier for local dipole switching, making the DG system highly susceptible to external stimuli like electric fields and promising for cryogenic energy storage [12, 31].

**High-performance cryogenic energy storge**

The unipolar *P*-*E* loops of the $Pb_{0.6}Sr_{0.4}ZrO_3$ films under various electric fields are further investigated at room temperature (300 K) (Supplementary Figs. 11-16), liquid-nitrogen temperature (77 K), and liquid-helium temperature (4 K), to evaluate their cryogenic energy-storage potential (Fig. 4a). Notably, very slim loops with low hysteresis are realized at ultra-high fields and maintained from room temperature down to 4 K due to the persistent glassy nature (Fig. 4a). Remarkably, $Pb_{0.6}Sr_{0.4}ZrO_3$ exhibits the highest $U_r$ of ~ 211 J/cm$^3$ at 77 K and maintains competitive $\eta$ over 88% down to 4 K even at an ultrahigh field of 9 MV/cm, delivering ultrahigh figure of merit $U_F$ (defined as $U_F = U_r/(1\text{-}\eta)$) ~1805 J/cm$^3$ at 77 K and ~1648 J/cm$^3$ at 4 K (Fig. 4b). The exceptionally high $U_r$ in $Pb_{0.6}Sr_{0.4}ZrO_3$ is attributed to two key factors. First, the lower leakage current and reduced dielectric loss effectively decreases energy dissipation in the form of *Joule* heat (Supplementary Fig. 12). Second, the disordered dipoles not only release internal strain originating from long-range lattice distortions but also moderate the structural phase transition, resulting in a high $E_B$ (Supplementary Figs. 5 and 14). The Weibull distribution analysis of the characteristic breakdown field ($E_B$) shows ultrahigh $E_B$ values of as high as ~ 8.50 MV/cm with Weibull modulus $\beta$ over 19.18 at 77 K, further confirming the high device-to-device reproducibility and reliability of the high-energy storage performance. Temperature-dependent $U_r$ and $\eta$ measured at a high field of 6 MV/cm demonstrate an ultrawide temperature stability window from 300 to 4 K, with variation of < 2.5% and < 3.7%, respectively (Fig. 4d and Supplementary Fig. 17). This remarkable thermal stability highlights one of the key advantages of this DG system, which is intrinsically unattainable in conventional RFEs.

Moreover, to evaluate the practical applicability in cryogenic conditions, the charge/discharge speed of the $Pb_{0.6}Sr_{0.4}ZrO_3$ capacitors, which is defined as the switching time of 90% of $P_{max}$ [32], was measured as a function of temperature (Fig. 4e). Under a 2 MV/cm pulse electric field, a charge/discharge speed of ~1.5 $\mu$s, with variation of < 3.1 % from room temperature down to 77 K, was observed, which is distinct from electrochemical energy-storage components that suffer gradually frozen kinetics at low temperatures [4]. Furthermore, under a 6 MV/cm electric field at 77 K, the energy storage performance of the $Pb_{0.6}Sr_{0.4}ZrO_3$ films maintains ultrahigh reliability and stability over $10^8$ charge/discharge cycles without noticeable degradation of energy density (variation < 4.5%) or efficiency (variation < 7.5%) (Fig. 4f), underscoring the competence of the DG system for practical cryogenic applications. Overall, the superior $U_r$ and $E_B$ position the DG-state PSZO films as competitive and promising candidates for dielectric capacitors compared to other dielectric materials (Fig. 4g) [8-11, 33-47]. Moreover, as for the cryogenic applications, the DG PSZO thin films represent remarkable advances over the previous efforts in the commercial biaxially oriented polypropylene (BOPP), multilayer ceramic capacitors (MLCC) and dielectric thin films (all limited to liquid-nitrogen temperature or above), representing, for the first time, the realization of robust operation at the extreme liquid-helium conditions combined with record-high dielectric energy storage density (Fig. 4h) [9, 10, 34, 36, 48-50].

**Conclusion**

In this study, we present a strategic approach to high-performance cryogenic energy storage by designing dipole-glass dielectrics with unit-cell-level polarization disorder near the AFE-PE phase boundary. We achieve such DG state in $Pb_{0.6}Sr_{0.4}ZrO_3$ thin films with remarkable dielectric energy-storage properties, including a record-high energy storage density $U_r$ up to 211 J/cm$^3$, excellent efficiency ($\eta$ > 88 %), robust stability, and more importantly, the first case reported to be functional at cryogenic temperatures down to 4 K. This DG state, achieved by disrupting the antipolar order of PZO with non-polar component, effectively complicates the dipole-dipole interactions and cancels the long-range ordering that typically leads to the formation of FE domains during cooling, thus stabilizes the unit-cell-level polar-disordering at cryogenic temperatures. This work not only sheds light on the mechanisms behind the enhanced (di)electric properties in DG materials but also offers a promising paradigm for achieving both high energy storage performance and low-temperature stability. This finding is expected to meet the growing demand for extreme-condition endeavors, paving the way for dielectric materials with superior performance and reliability in cryogenic applications.

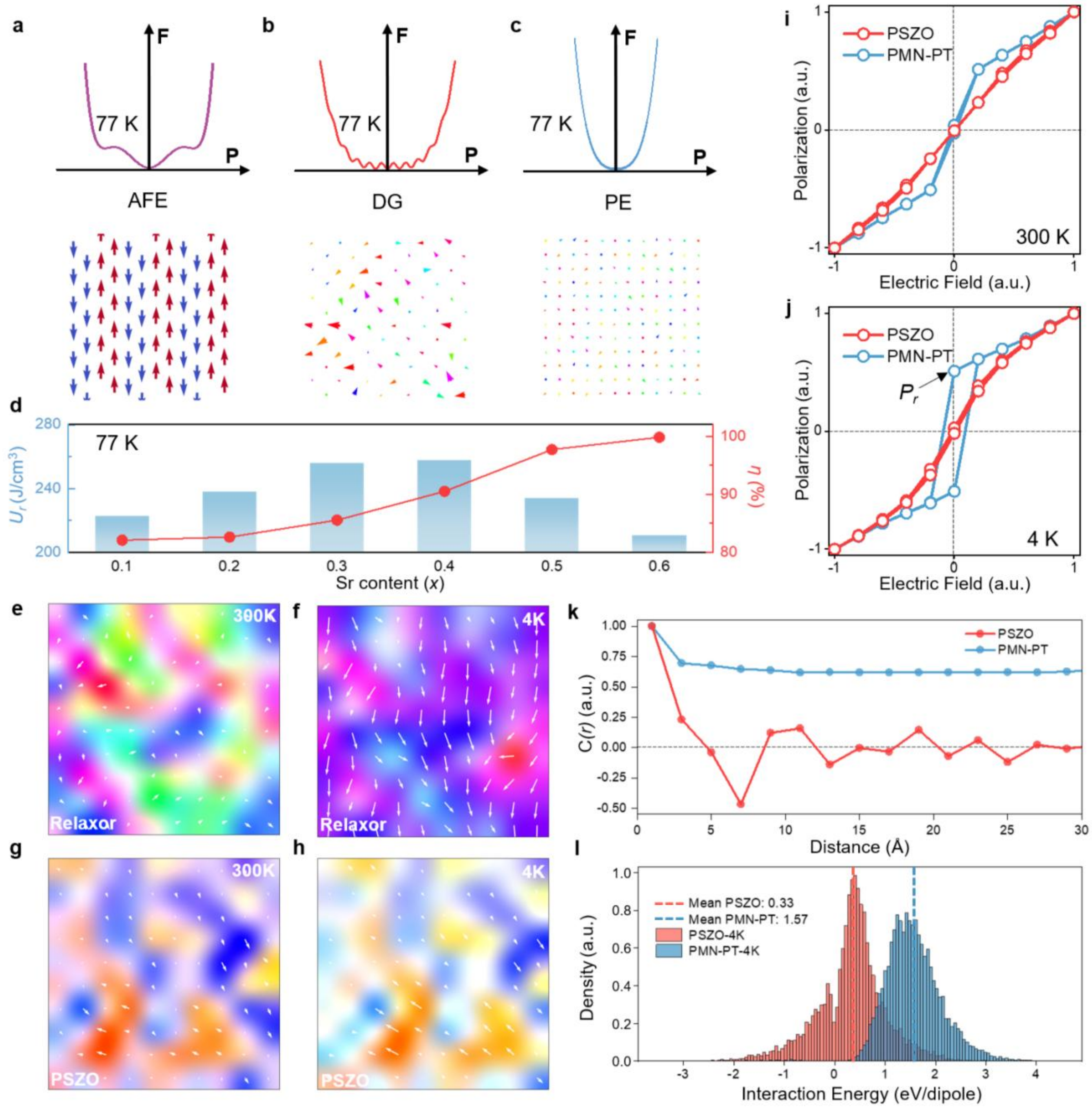

**Fig. 1 Simulations of dipole structures and energy-storage performances of $Pb_{1-x}Sr_xZrO_3$ thin films.** **a-c,** Statistic cation displacement distribution and free energy landscapes of (**a**) $PbZrO_3$, (**b**) $Pb_{0.6}Sr_{0.4}ZrO_3$ and (**c**) $SrZrO_3$ at a cryogenic temperature of 77 K. **d,** Energy storage density ($U_r$) and efficiency ($\eta$) for $Pb_{1-x}Sr_xZrO_3$ during the transition from AFE to DG states and to PE state. **e-h,** Dipole moment mappings of the typical RFE 70PMN-30PT at (**e**) 300 K and (**f**) 4 K, and the DG $Pb_{0.6}Sr_{0.4}ZrO_3$ at (**g**) 300 K and (**h**) 4 K, respectively. The brightness of background color indicates the magnitude of polarization. **i-j,** *P-E* loops of DG state $Pb_{0.6}Sr_{0.4}ZrO_3$ and RFE 70PMN-30PT at **(i)** 300 K and **(j)** 4 K. **k,** Spatial dipole-dipole correlation function of RFE 70PMN-30PT and DG $Pb_{0.6}Sr_{0.4}ZrO_3$. **l,** Distribution of dipole-dipole interaction energies of RFE 70PMN-30PT and DG $Pb_{0.6}Sr_{0.4}ZrO_3$.

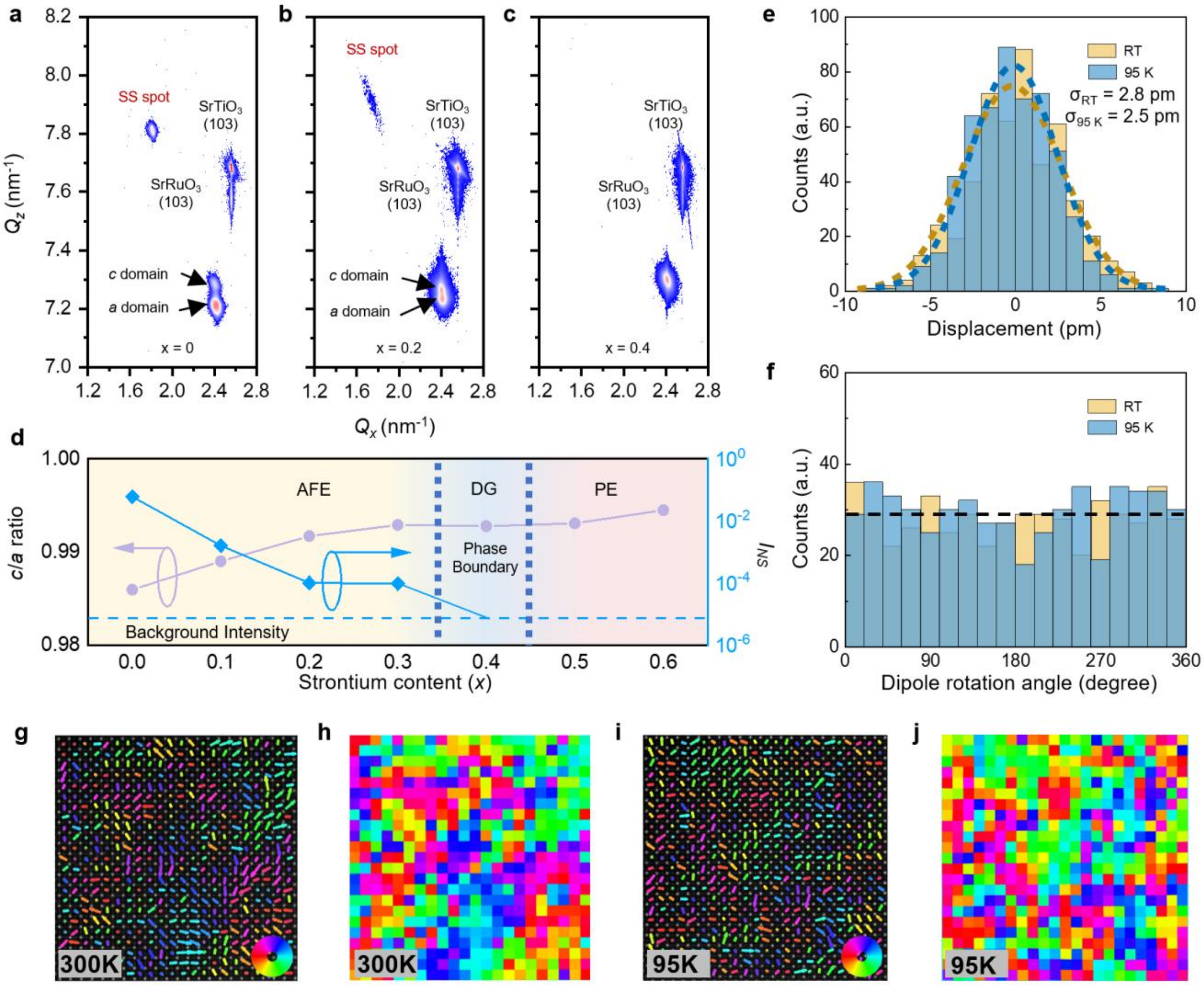

**Fig. 2 Structural evolution and the local dipole distributions in $Pb_{1-x}Sr_xZrO_3$ thin films. a-c,** 103-diffraction reflection of the films and substrates for (001)-oriented $Pb_{1-x}Sr_xZrO_3$ ($x$ = 0, 0.2, 0.4) heterostructures; the *a* domain and *c* domain correspond to the $[120]_o$ and $[001]_o$ directions, respectively, of the orthorhombic axis pointing towards the out-of-plane direction. **d,** *c*/*a* ratio and $I_{NS}$ of PSZO with respect to Sr substitution level *x*, where $I_{NS}$ is the normalized intensity of SS spot calculated as $I_{SS}/I_{(103)}$. **e,** Magnitude distribution of the A-site cation displacements projected along [101] at room temperature and 95 K. The orange and bule dashed lines in (**e**) are the Gaussian fitting of the histograms with the corresponding standard deviation labeled as $\sigma_{RT}$ and $\sigma_{95K}$, respectively. **f,** angle distribution of displacement directions at room temperature and 95 K. **g**, A-site cation displacement map and **h,** orientation map of dielectric dipoles in $Pb_{0.6}Sr_{0.4}ZrO_3$ films at room temperature. **i,** A-site cation displacement map and **j,** orientation map of dielectric dipoles in $Pb_{0.6}Sr_{0.4}ZrO_3$ films at 95 K.

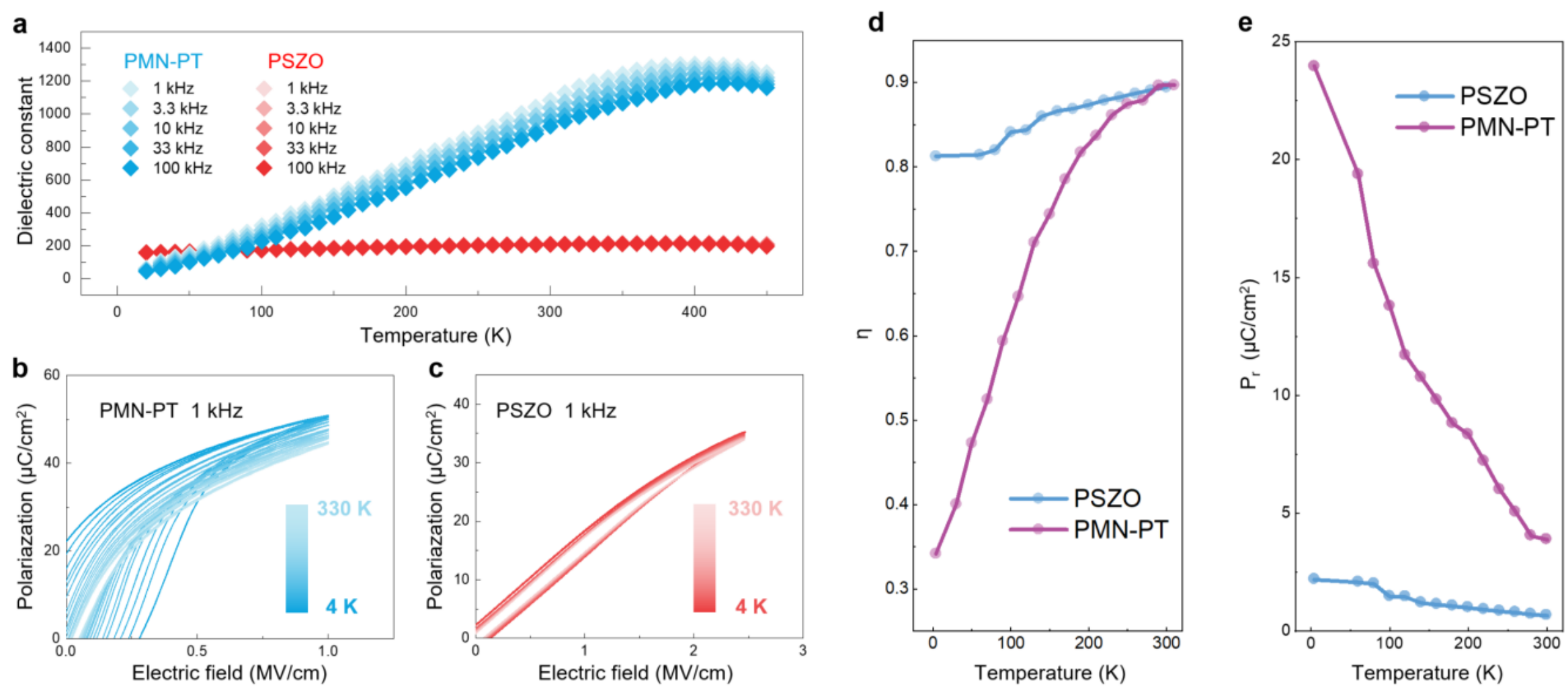

**Fig. 3 Temperature-driven electrical behavior evolution of DG $Pb_{0.6}Sr_{0.4}ZrO_3$ and relaxor PMNPT thin films. a,** Temperature-spectrum of dielectric permittivity for $Pb_{0.6}Sr_{0.4}ZrO_3$ and PMNPT. **b,** Hysteresis of relaxor PMN-PT thin films from 4 K to 330 K. **c,** Hysteresis of DG-state $Pb_{0.6}Sr_{0.4}ZrO_3$ thin films from 4 K to 330 K. **d,** Temperature-dependent efficiency and **e,** Temperature-dependent remanent polarization of PMN-PT and $Pb_{0.6}Sr_{0.4}ZrO_3$ from (**b**) and (**c**), respectively.

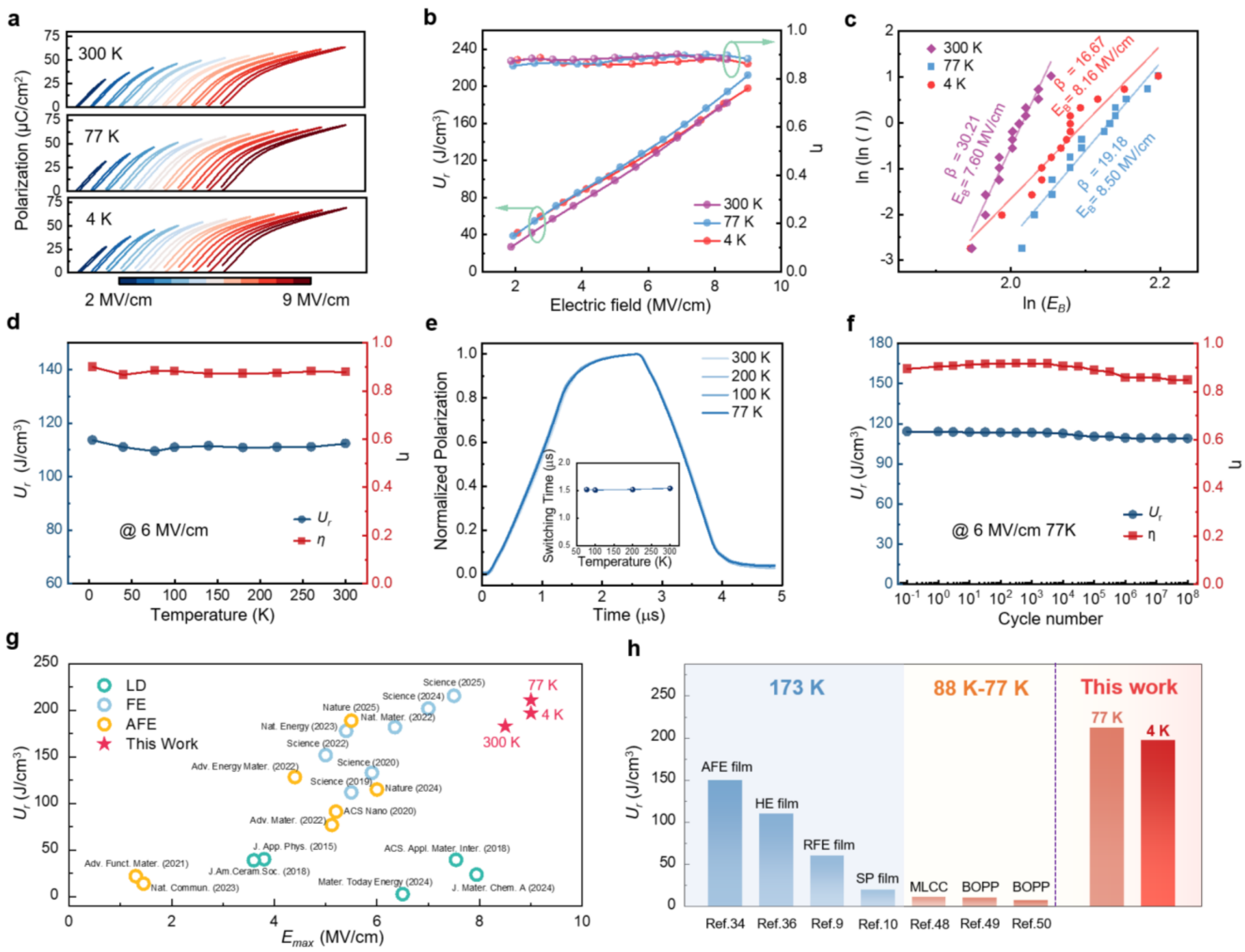

**Fig. 4 Dielectric energy storage properties of the DG-state films at cryogenic temperatures. a,** High electric-field driven unipolar P-E loop of DG-state $Pb_{0.6}Sr_{0.4}ZrO_3$ thin films under various electric fields at room temperature (upper panel), 77 K (middle panel), and 4 K (bottom panel). **b,** $U_r$ and $\eta$ at room temperature, 77 K and 4 K of the $Pb_{0.6}Sr_{0.4}ZrO_3$ thin films with respect to electric field. **c,** Weibull distribution of dielectric breakdown field of the $Pb_{0.6}Sr_{0.4}ZrO_3$ thin films at room temperature, 77 K and 4 K. **d,** Temperature dependent energy-storage performance of the DG-state films at an electric field of 6 MV/cm. **e,** Switching-dynamic study of the $Pb_{0.6}Sr_{0.4}ZrO_3$ thin films. The inset shows the switching time change with respect to temperature from 300K to 77 K. **f,** Energy storage performance of the $Pb_{0.6}Sr_{0.4}ZrO_3$ films at an electric field of 6 MV/cm at 77 K with respect to charging-discharging cycling numbers. **g,** Comparison of the energy-storage performance among the state-of-the-art dielectric materials including LD, FE and AFE; all data from the references are collected at room temperature [8-11, 33-47]. **h,** Comparison of the energy-storage density among commercial and recently reported dielectric capacitors, including BOPP, MLCC, superparaelectric (SP), relaxor ferroelectric (RFE), high-entropy (HE), and AFE thin films at cryogenic conditions [9, 10, 34, 36, 48-50].

## Method

**Molecular dynamic simulations.** Large-scale molecular dynamics simulations were conducted using LAMMPS with the Universal Perovskite Oxides Interatomic Potential (UniPero) [51]. An isobaric-isothermal (*NPT*) ensemble was employed, with temperature and pressure regulated by the Nosé-Hoover thermostat and Parrinello-Rahman barostat, respectively. The equations of motion were integrated using a timestep of 2 fs. An external electric field was applied using the force method, where the force induced by the electric field is given by $\boldsymbol{F} = \boldsymbol{Z}E$; $\boldsymbol{Z}$ represents the Born effective charge, and $E$ denotes the external electric field. To simulate the hysteresis loop, the electric field was applied incrementally in steps of 1 MV/cm, with the system being equilibrated for 100 ps at each step. The total energy storage density ($U_{\mathrm{total}}$) was calculated by integrating the polarization with respect to the electric field during the charging (field-increasing) phase, represented by the area between loading curve and the polarization axis: $U_{\mathrm{total}} = \int_0^{P_{max}} E\, dP$. Similarly, the recoverable energy density ($U_{\mathrm{r}}$), which is the energy released during the discharging (field-decreasing) phase, was calculated by integrating the area between the unloading curve and the polarization axis: $U_{\mathrm{r}} = \int_{P_r}^{P_{max}} E\, dP$. The energy loss ($U_{\mathrm{loss}}$) during one cycle is the difference between the $U_{total}$ and $U_r$, corresponding to the area enclosed by the hysteresis loop. Consequently, the energy storage efficiency ($\eta$) was determined as $\eta = \frac{U_r}{U_{total}} \times 100\%$. For the simulation of $Pb_{1-x}Sr_xZrO_3$(PSZO), a supercell size of $6\sqrt{2}\times10\sqrt{2}\times12$ was utilized, containing 7200 atoms. The simulation of (1-*x*)$Pb(Mg_{1/3}Nb_{2/3})O_3$-*x*$PbTiO_3$ (PMN-PT) employed a 10×10×10 supercell, housing 5000 atoms. The instantaneous local polarization $P_u(t)$ of each unit cell is calculated using the formula:

$$P_u(t) = \frac{1}{V_u}[\frac{1}{8}Z_A^*\sum_{i=1}^{8} r_{A,i}(t) + Z_B^* r_B(t) + \frac{1}{2}Z_O^*\sum_{i=1}^{6} r_{O,i}(t)]$$

Where $V_u$ is the volume of the unit cell. $Z_A^*, Z_B^*$ and $Z_O^*$ are born effective charges of A-site cation, B-site cation and oxygen ion, respectively. $r_{A,i}(t), r_B(t)$ and $r_{O,i}(t)$ are instantaneous atomic positions. During the simulation, the in-plane lattices were confined as constant as the substrate clamping.

**Dipole-Dipole interaction energy.** To quantify the local dipole-dipole interactions governing the ferroelectric behavior in PSZO and PMN-PT, we employed a computational method to probe the local effective field experienced by each A-site dipole. For each material, 10 unique supercells were constructed with a randomly generated distribution of cations to model the chemical heterogeneity. The systems were fully relaxed at a low temperature of 4 K for 50 ps, allowing both the lattice parameters and atomic positions to settle into their minimum energy state, ensuring the local dipole arrangement for each unique chemical ordering is robustly determined. Following relaxation, the effective interaction for each A-site dipole was quantified by calculating the energy cost, $\Delta E$, required to invert its orientation against the local field generated by its neighbors. This was achieved by systematically iterating through every A-site dipole, reversing its atomic displacement vector, and calculating the resultant change in the system's potential energy. The nature and strength of the interaction were then directly inferred from this energy cost. The magnitude of the interaction energy is proportional to ΔE. The sign of the interaction—positive for a ferroelectric (parallel-aligning) preference and negative for an anti-ferroelectric (anti-parallel-aligning) preference—was determined by the alignment of the central dipole with the net polarization of its nearest neighbors. This method yields a statistical distribution of interaction energies for each material, allowing for a direct quantitative analysis of the competing forces that govern the system's propensity for long-range order versus a disordered, glassy state. To quantify the spatial extent of polar correlations, the dipole-dipole spatial autocorrelation function, C(r), was calculated, which is defined as $C(r) = \langle P_i \cdot P_j \rangle_r$, representing the average dot product of normalized dipole moments for all pairs $(i, j)$ separated by a distance $r$.

**Fabrication of $Pb_{1-x}Sr_xZrO_3$ epitaxial thin films.** High-quality $Pb_{1-x}Sr_xZrO_3/SrRuO_3$ (PSZO/SRO) heterostructure was grown on (001)-oriented $SrTiO_3$ (STO) single crystal substrate by pulsed laser deposition (PLD) (Arrayed Materials RP-B) using a 248-nm KrF excimer laser. SRO was first deposited on the substrate and was used as the bottom electrode. The SRO layer was deposited at a substrate temperature of $T_g$ = 680 ºC and a dynamic oxygen pressure $P_{O2}$ = 100 mTorr, with a laser fluence of 1.0 $J/cm^2$ on the target. Subsequently, PSZO films were deposited at a lower substrate temperature of 590 ºC and a dynamic oxygen pressure of 80 mTorr, with a laser fluence of 1.6 $J/cm^2$. After growth, the samples were cooled down to room temperature at a rate of 5 °C $min^{-1}$ under an oxygen pressure of 700 Torr. Top electrode pattern of 50 μm and 25 μm diameter circle arrays was transferred to the top of PSZO films by photolithography with light wavelength λ = 385 nm and resolution μ ≤ 1 μm. Top electrode Pt was then deposited by magnetron sputtering (Arrayed Materials RS-M). After sputtering, the sample was dipped into acetone and clean with ultrasonic washer for 30 s to remove the photoresist.

**Multislice electron ptychography reconstruction.** The 4D-STEM datasets for multislice electron ptychography reconstruction were acquired on an aberration-corrected JEM-ARM200 transmission electron microscope equipped with a Dectris Arina detector. The electron probe has a convergence semi-angle of 28.2 mrad and current of approximately 220 pA. Each 4D-STEM dataset has 200×200 diffraction patterns with a step size of about 0.498 Å and a dwell time of 34 μs. Thus, the acquiring time is 1.36 s and illumination dose was 1.9 × 105 e−/Å$^2$. Each of the original diffraction patterns is 192 × 192 in size and was further binned to 96 × 96 before the ptychographic reconstruction. Cryogenic experiments were performed using a double-tilt liquid-nitrogen cooling holder (JEOL EM-31660) maintained at approximately 95 K. To ensure thermal equilibrium and measurement stability, data acquisition was initiated after a 3-hour temperature stabilization period. Multislice electron ptychographic reconstruction used the least-squares maximum-likelihood algorithms developed earlier for X-ray ptychography [52] and later optimized for electron microscopy [53]. The sample thickness is approximately 14 nm, and the reconstruction used a slice thickness of 0.8 nm and four probe modes based on mixed-state model [54]. The code for multislice electron ptychography is adapted from the previously published versions in Zenodo [55].

**Electrical measurements.** The room-temperature electrical measurements were performed on a Semishare E4 high-precision probe station, while the low-temperature *P-E* loop measurement was completed at Lake Shore PS-100 tabletop cryogenic probe station. TF3000 analyzer (aixACCT) with additional voltage amplifier ensures the high-field measurements of energy storage performance. The *P-E* loops of PSZO films were characterized under a 5-kHz AC electric field for both unipolar and bipolar conditions. The applied voltage was set to increase gradually with step width of ~2 V until the electrode breakdown occurs. The fatigue performance was conducted at 77 K with a 100-kHz AC cycling field ~ 6 MV/cm and a 5-kHz AC testing field ~6 MV/cm. The switching dynamic was characterized at 77 K, 100 K, 200 K, and 300 K, respectively, with a pulse electric field up to 2 MV/cm and rise time of 1.5 μs. The frequency spectrum of dielectric constant was collected with an E4980A LCR meter (Agilent Technologies), and the test voltage was set to be ~500 mV. The leakage current was collected with a 4200A-SCS semiconductor analyzer (Keithley), and each point was averagely recorded with staying time of 1 s.

**Data availability**

All data supporting the findings of this study are available within the Article and Supplementary Information.

**Code availability**

The code used to calculate the results shown in this work is available from the corresponding authors upon reasonable request.

**Acknowledgements**

This project was financially supported by National Natural Science Foundation of China (Grant Nos. 52525209, 52372105 and 92477129), Guangdong Basic and Applied Basic Research Foundation (Grant No. 2024B1515120010), and Shenzhen Science and Technology Program (Grant No. KQTD20200820113045083). Z.H.C. has been supported by State Key Laboratory of Precision Welding & Joining of Materials and Structures (Grant No. 24-Z-13) and “the Fundamental Research Funds for the Central Universities” (Grant No. 2024FRFK03012). Z.C. acknowledge funding support from the National Key Research and Development Program of China (MOST) (Grant No. 2023YFA1406400) and National Natural Science Foundation of China (Grant No. 52273227).

**Author Contributions**

Z.H.C. and Y.Y.S. conceived of and designed the research. Z.H.C. supervised the project. Y.J.L., Y.Y.S., and J.K. grow the films, measured X-ray diffraction and electric performance with the assistance of J.X.L., T.F.Z., J.L.W, C.Z., H.X, Z.Q.R., D.L. J.K.Q. and C.Y.X.. D.N.L. and S.L. performed first-principles calculations. C.S.C., H.C.Y., Y.Z., Y.L.T., and Z.C. performed STEM and MEP measurements. Y.Y.S., Y.J.L., D.N.L., C.S.C., H.P., S.D., F.L., Z.C., S.L., and Z.H.C. analyzed the data. Y.Y.S., Y.J.L., H.P., S.L., and Z.H.C. wrote the manuscript. All authors discussed and commented on the manuscript.

**Competing interests**

The authors declare no competing financial interests.

**Additional information**

Supplementary Information is available for this paper. Correspondence and requests for material should be addressed to Zuhuang Chen (zuhuang@hit.edu.cn).